\documentclass[aps,twocolumn,prb,10pt,superscript,floatfix,superscriptaddress,showpacs,footinbib]{revtex4-2}

\usepackage{amssymb}

\usepackage{graphicx}
\usepackage{amsmath}
\usepackage{amsmath,bm}
\usepackage{blindtext, xcolor}
\usepackage{xcolor}
\usepackage{hyperref}
\usepackage{nicefrac}
\hypersetup{colorlinks,citecolor=blue,linkcolor=blue,urlcolor=blue}

\newcommand{\dlangle}{\langle\langle}

\newcommand{\drangle}{\rangle\rangle}

\begin{document}
\title{Kondo effect in a two-dimensional electron gas in the Persistent Spin Helix regime}

\author{T.~O.~Puel}
\affiliation{Department of Physics and Astronomy, University of Iowa, IA 52242, USA}
\author{M.~A.~Manya}
\affiliation{Universidad Tecnologica del Per\'u, Lima, Perú}
\affiliation{Instituto de F\'isica, Universidade Federal Fluminense, Niter\'oi, RJ 24210-340, Brazil}
\author{G.~S.~Diniz}
\affiliation{Curso de F\'isica, Universidade Federal de Jata\'i, Jata\'i, GO 75801-615, Brazil.}
\affiliation{Instituto de F\'isica, Universidade Federal de Catal\~ao, Catal\~ao, GO 75705-220, Brazil.}
\author{E.~Vernek}
\affiliation{Department of Physics and Astronomy, Ohio University, Athens, OH 45701-2979, USA}
\affiliation{Instituto de F\'isica,  Universidade Federal  de Uberl\^andia, Uberl\^andia, MG 38400-902, Brazil}
\author{G.~B.~Martins}
\affiliation{Instituto de F\'isica,  Universidade Federal  de Uberl\^andia, Uberl\^andia, MG 38400-902, Brazil}
\date{\today}

\begin{abstract}
The Kondo effect arises from many-body interactions between localized magnetic impurities and conduction electrons, affecting electronic properties at low temperatures. In this study, we investigate the Kondo effect within a two-dimensional electron gas subjected to strong spin-orbit coupling in and out of the persistent spin helix regime, a state characterized by a long spin lifetime due to SU(2) symmetry recovery. 
Using the numerical renormalization group approach, we systematically analyze the influence of spin-orbit coupling strength and the orientation of an external magnetic field on the spectral properties of the impurity. 
Our findings reveal an entrancing interplay between spin-orbit coupling and the magnetic field, leading to key phenomena such as splitting of the hybridization function, asymmetry in the spectral function of the impurity, and significant tunability of the Kondo temperature due to spin orbit.
These results provide valuable insights into the delicate balance between spin-orbit and external magnetic field effects in quantum impurity systems, contributing to a deeper understanding of spintronics and quantum manipulation in low-dimensional materials.
\end{abstract}

\maketitle

\section{Introduction}

The unexpected minimum in the resistance curve of gold wires that was observed
as the temperature was reduced became an intriguing puzzle in the work of W. de Haas
\emph{et al.}~\cite{Dehaas1934} in 1934. This long-standing problem was solved only 30 years later by
Jun Kondo in his groundbreaking paper~\cite{Kondo1964}, where he applied 
perturbation theory to explain the violation of Matthiessen's rule, i.e., linear decay 
of resistivity with temperature. To explain this violation, Kondo suggested that a 
many-body electron scattering mechanism occurs at a specific low-temperature scale. 
Using perturbation theory, he showed that, as the temperature was lowered, a localized 
magnetic moment (originating from magnetic impurities in the metallic sample) 
should interact with the conduction band electrons, thus forming a spin singlet state, 
which causes a logarithmic contribution to the low-temperature resistivity~\cite{Kondo1964}. 
When Kondo was able to explain the experimental observation in the resistance minimum, 
there were already several other experiments (on related metals) observing the characteristic 
temperature $T_{K}$, later dubbed the `Kondo temperature'~\cite{Hewson1993}, below which 
 the logarithm increase in the resistance emerges. 

Currently, the so-called Kondo effect is still a subject of intense research. 
From a theoretical standpoint, most of the advances were possible thanks to the
impurity solvers developed from the late 70s on~\cite{Wilson1975,Krishna-murthy1980,Bulla2008}. 
For instance, the Kondo effect has been extensively studied in several different materials, 
with distinct spatial confinement~\cite{Nygard2000,science.281.5376.540,PhysRevLett.100.026807,PhysRevB.92.121109} 
and magnetic impurities~\cite{Otte2008,PhysRevB.78.224404}. Moreover, structural disorder~\cite{PhysRevB.90.201101} 
and different external conditions, such as  applied magnetic fields, can play a fundamental role~\cite{PhysRevLett.85.1504,Zhang2013,PhysRevB.87.205313}. 
Lastly, the Kondo effect in the presence of spin-orbit coupling (SOC) has also
been widely studied in the last few years~\cite{PhysRevB.80.041302,PhysRevB.94.125115,Zitko2011,PhysRevLett.108.046601,Martins2020}.

Indeed, SOC, an essential interaction that explains the atomic electronic structure, 
gained much attention in the semiconductor field after G. Dresselhaus' pioneering work 
on graphite structures~\cite{Dresselhaus1955}. 
A few years after Dresselhaus' work, theoretical predictions by Bychkov and Rashba
demonstrated that induced electric fields, generated by inversion asymmetry in
two-dimensional (2D) systems, were responsible for an effective SOC~\cite{Bychkov1984}, later
dubbed Rashba SOC. The theoretical predictions were subsequently 
observed, mainly because of the fast development of state-of-the-art growth and 
characterization techniques, which in turn enabled the manipulation of the 
spin degree of freedom \cite{science.1065389}. Interestingly, proposals of spin manipulation by means of 
SOC in 2D structures were able to demonstrate intriguing phenomena. For instance, 
by virtue of Rashba SOC, an all-electric field effect transistor has been proposed~\cite{DattaDas1990}, 
which generated numerous theoretical and experimental studies, especially because 
of its fundamental role in the development of spin-based 
electronics~\cite{RevModPhys.76.323,PhysRevApplied.4.047001}. 
Nowadays, Rashba SOC became a dominant ingredient in the application of topology to condensed matter, 
with major examples on quantum spin transport~\cite{Bercioux_2015}, 
topological band structure \cite{Zhou2014}, and quantum spin Hall effect~\cite{JPSJ.77.031007}.

Interestingly, by exploring the manipulation of the emergent Rashba and 
Dresselhaus SOC strengths in 2D materials, Schliemann \emph{et al.} 
demonstrated that for equal Rashba and Dresselhaus SOC parameters, 
a momentum-independent spin texture appeared~\cite{Schliemann2003}. 
This intriguing result was further theoretically studied by Bernervig \emph{et al.}~\cite{Bernevig2006}, 
who demonstrated the emergence of SU(2)-symmetry-recovery for the specific condition 
of equal strength of Rashba and Dresselhaus SOC. As a direct consequence of this result, a 
long-lived helical spin excitation is expected for a specific momentum orientation, 
with a so-called persistent spin helix (PSH) state. 
Experimental observation of that state was later made in GaAs 
quantum wells, where, by controlling doping and the width of the 
quantum well, the Dresselhaus and Rashba SOC parameters could 
be precisely engineered to the required regime~\cite{Koralek2009}. 
That experimental success has led to various proposals and 
attempts to observe persistent spin textures in different 
materials, setups, and systems, like ferroelectrics~\cite{Lee2020,Huinan2024}, 
through laser-assisted SOC control~\cite{Li2022,Xue2023}, electro-optical 
SOC control~\cite{Xue2024}, through application of uniaxial stress~\cite{Kashikar2023}, 
and also in recently discovered 2D materials~\cite{Ulil2021,Sasmito2021,
Ji2022,Absor2022,Guo2023}. 
 
Despite numerous studies of systems in the PSH state~\cite{Jaroslav2009,Slipko2011,Walser2012,Dollinger2014,Kunihashi2016,
Kozulin2017,Kohda2017,Schliemann2017,Fu2018,Karimi2018,Passmann2019,Lu2020,Liu2020,Tkach2022,
Saito2022,Ishihara2022,Zhao2023,Lu2023}, 
little attention has been given to the Kondo effect when 
the host is in the PSH regime~\cite{Johannesson2011}. In this paper, we aim at bridging this 
important gap through numerical investigation of a magnetic impurity coupled to a two-dimensional 
electron gas (2DEG) in the PSH regime. To perform the calculations, we use 
the numerical renormalization group (NRG)~\cite{Bulla2008,PhysRevB.79.085106}. 
Our results reveal that SOC induces a splitting in the spectral function, leading to a suppression of the Kondo temperature. However, we also demonstrate the recovery of this state under the PSH condition. Additionally, we investigate the vectorial nature of the spectral function in the presence of a magnetic field, uncovering an asymmetry in the double-peak structure that depends on the field orientation.

This paper is organized as follows: In Sec.~\ref{sec: hamiltonian}  we present the Hamiltonian of the system  and detail how we use the NRG approach to address the Kondo regime.
In Sec.~\ref{sec-NRG}, we present the results, divided into three subsections:
Sec.~\ref{zero-field} NRG results without a magnetic field, analyzing the spin-orbit effects;
Sec.~\ref{sec: mag field} the introduction of a magnetic field and the analysis of its effect on the Kondo peak, 
as its strength is varied; and Sec.~\ref{sec: mag field orientation} the variation of the magnetic field direction, highlighting differences between distinct spin-orbit couplings, with a particular focus 
on the PSH regime, i.e., equal Rashba and Dresselhaus SOC.
Finally, our concluding remarks are presented in Sec.~\ref{sec: conclusions}.

\section{Model Hamiltonian and Hybridization Function}
\label{sec: hamiltonian}
\subsection{Model Hamiltonian}

\begin{figure}[h!]
\includegraphics[width=1.0\columnwidth]{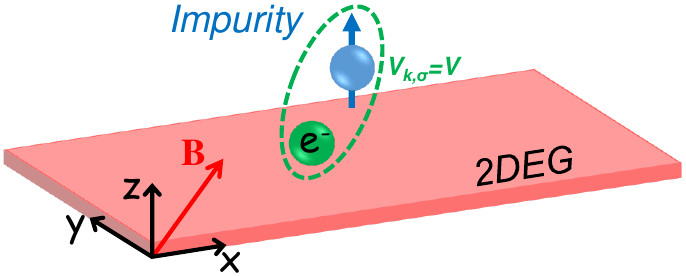} 
\caption{Schematic representation of a magnetic impurity coupled 
	(through a $k$-independent strength) to a 2DEG with strong 
	SOC. An external magnetic field $\mathbf{B}$, applied in 
	an arbitrary direction, is used to probe  
	the effective SOC-generated magnetic field $\mathbf{B_{SO}}$ 
	(induced by the combination of Rashba and Dresselhaus SOC), 
	through its effect over the Kondo state.
\label{fig1}}
\end{figure}

We consider a system formed by a magnetic impurity embedded in a two-dimensional-electron gas with strong spin-orbit coupling and subjected to an external magnetic field, $\mathbf{B}$, as shown in Fig.~\ref{fig1}. 
The  total Hamiltonian of the system can be expressed concisely as
\begin{eqnarray}
H = H_{\rm 2DEG} + H_{\rm imp} + H_{\rm hyb}.
\label{eq:Hamilt0}
\end{eqnarray}
Here $H_{\rm 2DEG}= H_{\rm 0} + H_{\rm SOC} + H_{\rm Z}$ describes conduction electrons, which comprises the kinetic energy ($H_{\rm 0}$), SOC ($H_{\rm SOC}$), the Zeeman effect caused by the external magnetic field  ($H_{Z}$). Explicitly, each of these terms can be written as,
\begin{eqnarray}
H_{\rm 0}&=& \sum_{\boldsymbol{k}} \varepsilon_{\boldsymbol{k}} ( c^\dagger_{\boldsymbol{k},\uparrow} c_{\boldsymbol{k},\uparrow} + c^\dagger_{\boldsymbol{k},\downarrow} c_{\boldsymbol{k},\downarrow} ), \\
H_{\rm SOC}&=& \sum_{\boldsymbol{k}} [\alpha(k_y + i k_x)+\beta( k_x + i k_ y)] c^\dagger_{\boldsymbol{k},\downarrow} c_{\boldsymbol{k},\uparrow} \nonumber \\
& & + \text{H.c.}, \\
H_{\rm Z}&=& g\mathbf{B}\cdot \boldsymbol{S}.
\label{eq:Hamilt1}
\end{eqnarray}
In these expressions, $c^\dagger_{\boldsymbol{k},\sigma}$ ($c_{\boldsymbol{k},\sigma}$) represents  the creation (annihilation) operator of an electron with momentum $\boldsymbol{k}$ and spin $\sigma$ within the 2DEG, $\varepsilon_{\boldsymbol{k}} = \hbar^2 (k_x^{2} + k_y^{2}) / 2m^{*}$, $\boldsymbol{k}=(k_x, k_y, 0)$ lies on the $xy$-plane. Finally, $\boldsymbol{S}$ is the total spin operator of the conduction electrons, and $\boldsymbol{B} = \mu_B (B_x, B_y, B_z)$ has units of energy.
The SOC Hamiltonian, $H_{\rm SOC}$, carries contributions both Rashba and Dresselhaus type of couplings, with strength $\alpha$ and $\beta$, respectively, and are induced by the lack of inversion symmetry~\cite{Silsbee_2004}.
$g$ is the electron's $g$-factor within the 2DEG. The Hamiltonian $H_{\rm 2DEG}$ can still be recast in the compact form~\cite{Kohda_2017}
\begin{eqnarray}
H_{\rm 2DEG}&=& H_{\rm 0}+g\mathbf{B_{tot}}\cdot\bm{S},
\label{eq:Hamilt1compact}
\end{eqnarray}
where $\mathbf{B_{tot}}=\mathbf{B} + \mathbf{B_{SO}}$ is 
the combined effect of the external field and the $k$-dependent SOC effective field given by $\mathbf{B_{SO}}=(\alpha k_{y} + \beta k_{x},-\alpha k_{x} - \beta k_{y}, 0)/g$. 
For convenience, hereon we will parametrize the spin-orbit term by $\gamma \geq 0$ (scaling the total SOC strength) and $0 \leq \theta \leq \pi/2$ (quantifying the contribution from each type of coupling), namely, 
\begin{equation}
 \alpha=\gamma\sin(\theta) \qquad \text{and} \qquad \beta=\gamma\cos(\theta).
 \label{eq: SOC strength with theta}
\end{equation} 
Clearly, $\gamma^2 = \alpha^2 + \beta^2$, and $\theta = 0$ and $\theta = \pi/2$ restricts the SOC to Dresselhaus  and Rashba only, respectively.

The second term in Eq.~\eqref{eq:Hamilt0} is given by,
\begin{eqnarray}
H_{\rm imp}&=& \varepsilon_{d} ( n_{\uparrow} + n_{\downarrow} ) + Un_{\uparrow}n_{\downarrow} + g_\text{imp}\mathbf{B}\cdot \boldsymbol{S}_\text{imp}, 
\label{eq:Hamilt2}
\end{eqnarray}
where $n_{\sigma}=d^\dagger_{\sigma}d_{\sigma}$ is the electron number operator, in which $d^\dagger_\sigma$ ($d_\sigma$) creates (annihilates) and electron with spin $\sigma = \, \uparrow, \downarrow$ in the impurity energy level $\epsilon_d$. In the last expression above, $g_\text{imp}$ represents the impurity g-factor, $\boldsymbol{S}_\text{imp}$ is the total spin, and $U$ is the Coulomb repulsion at the impurity.
Finally, the hybridization between the localized impurity and the 2DEG is given by
\begin{eqnarray}
H_{\rm hyb} = \sum_{\boldsymbol{k}} V_{\boldsymbol{k}} (d^\dag_{\uparrow}c_{\boldsymbol{k},\uparrow} +d^\dag_{\downarrow}c_{\boldsymbol{k},\downarrow} ) + \text{H.c.},
\label{eq:Hamilt3}
\end{eqnarray}
where we used the approximation, $V_{\boldsymbol{k}}=V$, i.e. a momentum- and spin-independent hybridization between the conduction and impurity orbitals, as discussed in Ref.~\cite{Krishna-murthy1980}.

\subsection{Hybridization Function}
To obtain the impurity spectral function within
the numerical-renormalization group (NRG)~\cite{Wilson1975}, we compute the impurity Green's function, that can be formally written in the impurity's spin basis as~\cite{PhysRevB.102.155114},
\begin{eqnarray}
\hat{G}_{\rm imp}(\omega)\!=\!\left[(\varepsilon_{d}-\omega)\sigma^{0}-\hat{\Sigma}^{(0)}(\omega)-\hat{\Sigma}^\text{(int)}(\omega)\right]^{-1}, 
\label{eq:Hybrid0}
\end{eqnarray}
where $\hat{\Sigma}^\text{(int)}(\omega)$ is the many-body self-energy.
As an input, we provide the non-interacting self-energy correction to the impurity after integrating out the 2DEG bath, that is, $\hat{\Sigma}^{(0)}(\omega)=\sum_{ \boldsymbol{k} }\hat{V} \hat{G}_\text{2DEG}( \boldsymbol{k},\omega)\hat{V}^{\dagger}$, with $\hat{V}=V\sigma^{0}$ and $\hat{G}_\text{2DEG}(\boldsymbol{k},\omega)=\left[\omega\sigma^{0} -H_\text{2DEG}(\boldsymbol{k})\right]^{-1}$, where $\sigma_0$ is the $2\times2$ identity matrix.
The $H_\text{2DEG}$ keeps the form of Eq. (\ref{eq:Hamilt1compact}), i.e., as written in the conduction-band-spin basis.
Because of the spin mixing, which is naturally arising from the SOC and the external magnetic field in the $xy$ plane, $\hat{\Sigma}^{(0)}(\omega)$ has off-diagonal elements (i.e., channel mixing, see Appendix~\ref{app-explicit-G} and \ref{app-band-polarization} for more details), such that the NRG requires the spectral representation of the hybridization [or hybridization function, $\hat{\Gamma} (\omega)$] to be characterized by the advanced and retarded self-energy components~\cite{nanjing,PhysRevB.95.035107}, written as
\begin{eqnarray}
\hat{\Gamma}(\omega)\!=\!\frac{1}{2i}\int\int 
	\left[\hat{\Sigma}^{(0)}( \boldsymbol{k},\omega-i\eta)-\hat{\Sigma}^{(0)}( \boldsymbol{k},\omega+i\eta)\right]d^{2}k, \nonumber \\ 
\label{eq:Hybrid1}
\end{eqnarray}
where $\eta \rightarrow 0^{+}$.
This $2\times2$ hybridization function matrix requires a non-conventional Wilson RG scheme~\cite{nanjing}.
Note that $\Gamma(0) = \pi V^2 \rho(\varepsilon_F)$, where $\rho(\omega) = -(1/\pi) \text{Im} [G_\text{2DEG} (\omega)]$.

\begin{figure}[h!]
\includegraphics[width=1.0\columnwidth]{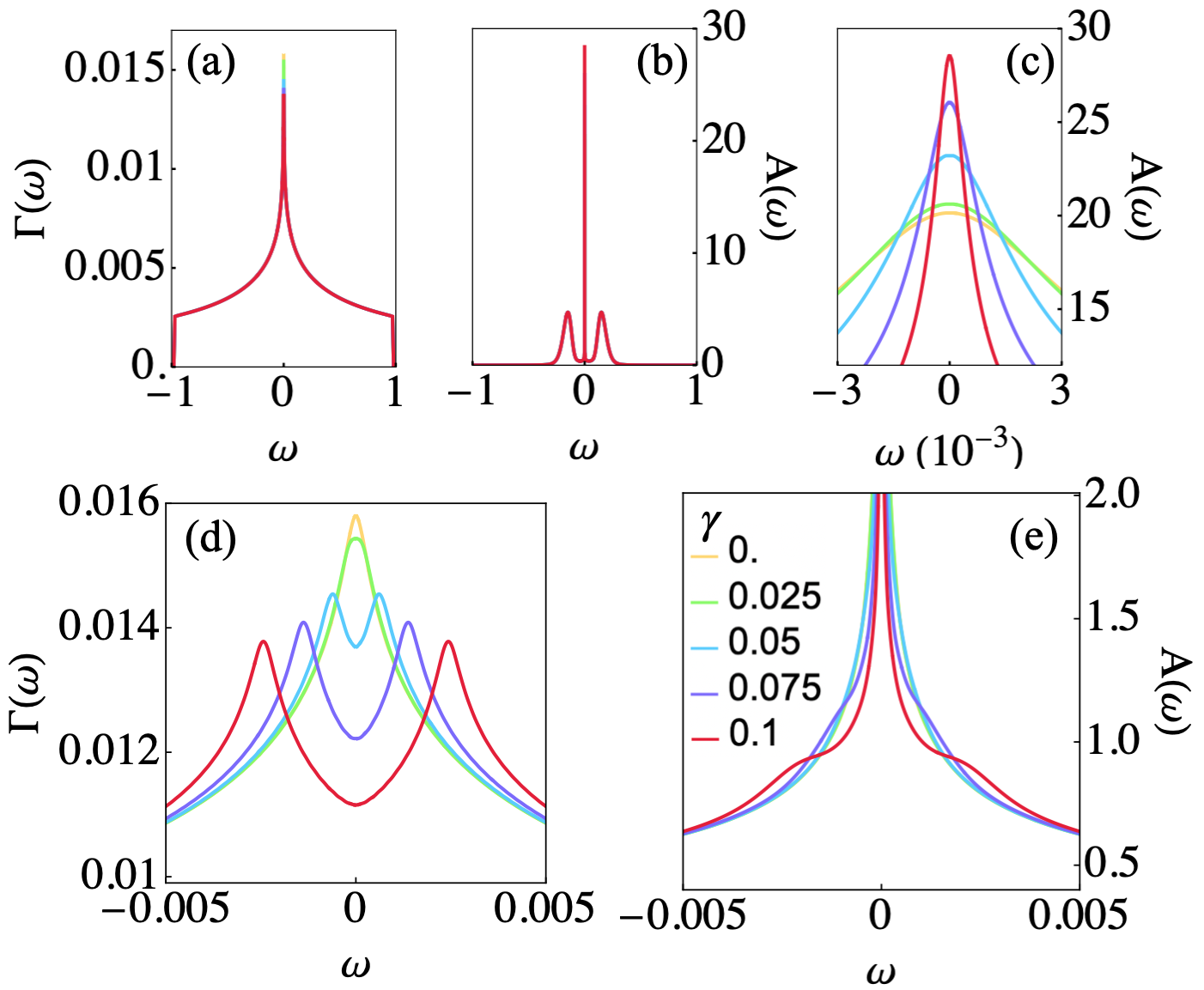} 
    \caption{(a) shows the hybridization function 
    $\Gamma(\omega) = \Gamma_\uparrow(\omega) = \Gamma_\downarrow(\omega)$, in absence of magnetic field,
    over the full range of band frequency $\omega$ for several SOC strengths $\gamma$, including only Rashba coupling; $\theta = \pi /2$. 
    All panels follow the legends in (e). 
    A zoom in on the hybridization function peak is shown in (d). 
    (b) shows the magnetic impurity's spectral function $A(\omega)$ for several SOC strengths $\gamma$.
    A zoom in on the Kondo peak is shown in (c).
    For sufficiently large value of SOC strength the Kondo peak develops shoulders in its basis, as shown in (e) and most prominent for $\gamma=0.1$ (red curve).}
    \label{fig2}
\end{figure}

\section{NRG results}\label{sec-NRG}
\subsection{Kondo dependence with SOC: zero magnetic field}
\label{zero-field}

For the results below, we have set $V = 0.05$, $U = 0.3$, and $\epsilon = -U/2$ (the particle-hole symmetric point). For further details, see Appendix \ref{app-method-details} and \ref{app-representative-parameters}. We begin our analysis by examining the Kondo effect as we change the strength of the spin-orbit coupling  in the absence of an external magnetic field. 
In this case, we compute the hybridization function, $\Gamma(\omega) = \Gamma_\uparrow(\omega) = \Gamma_\downarrow(\omega)$, which serves as the input to the NRG method, as discussed above. 
For now, only Rashba type of SOC is included in the conduction band ($\theta = \pi/2$).
In Fig.~\ref{fig2}(a), $\Gamma (\omega)$ displays the characteristic shape expected for a 2DEG~\cite{Economou}. 
However, upon closer inspection of the peak structure, we observe a peak splitting, as SOC is introduced~\cite{PhysRevLett.128.027701}, as shown in Fig.~\ref{fig2}(d). The peak splitting is approximately on the order of the SOC strength.
Notice that the value of $\Gamma(0)$, which is important for the Kondo effect, decreases monotonically as $\gamma$ increases. 
Specifically, $\Gamma(0)$ takes on the values $\Gamma(0)=0.0158,~0.0154,~0.0137,~0.0122,~0.0112$ as $\gamma$ progressively increases within the values shown in Fig.~\ref{fig2}.

Next, we analyze the spectral function of the magnetic impurity, $A(\omega)$, obtained by the NRG calculations for various SOC strengths, $\gamma$.
Figure~\ref{fig2}(b) shows a sharp Kondo peak at $\omega = 0$ is observed between the Hubbard peaks.
A closer look at the Kondo peak structure in Fig.~\ref{fig2}(c) reveals a narrowing of the peak and an increase in its height as the SOC strength increases.
Since the Kondo temperature ($T_K$) is proportional to the half-width at half-maximum of this peak, we anticipate a lower $T_K$ for higher SOC strengths.
Finally, at the base of the Kondo peak, we observe the formation of ``shoulders'' for sufficiently large SOC values, as noticeable in the $\gamma = 0.075$ and $\gamma = 0.1$ curves in Fig.~\ref{fig2}(e).
We associate these shoulders with the splitting in $\Gamma(\omega)$, shown in Fig.~\ref{fig2}(d).

\begin{figure}[th!]
\includegraphics[width=0.95\columnwidth]{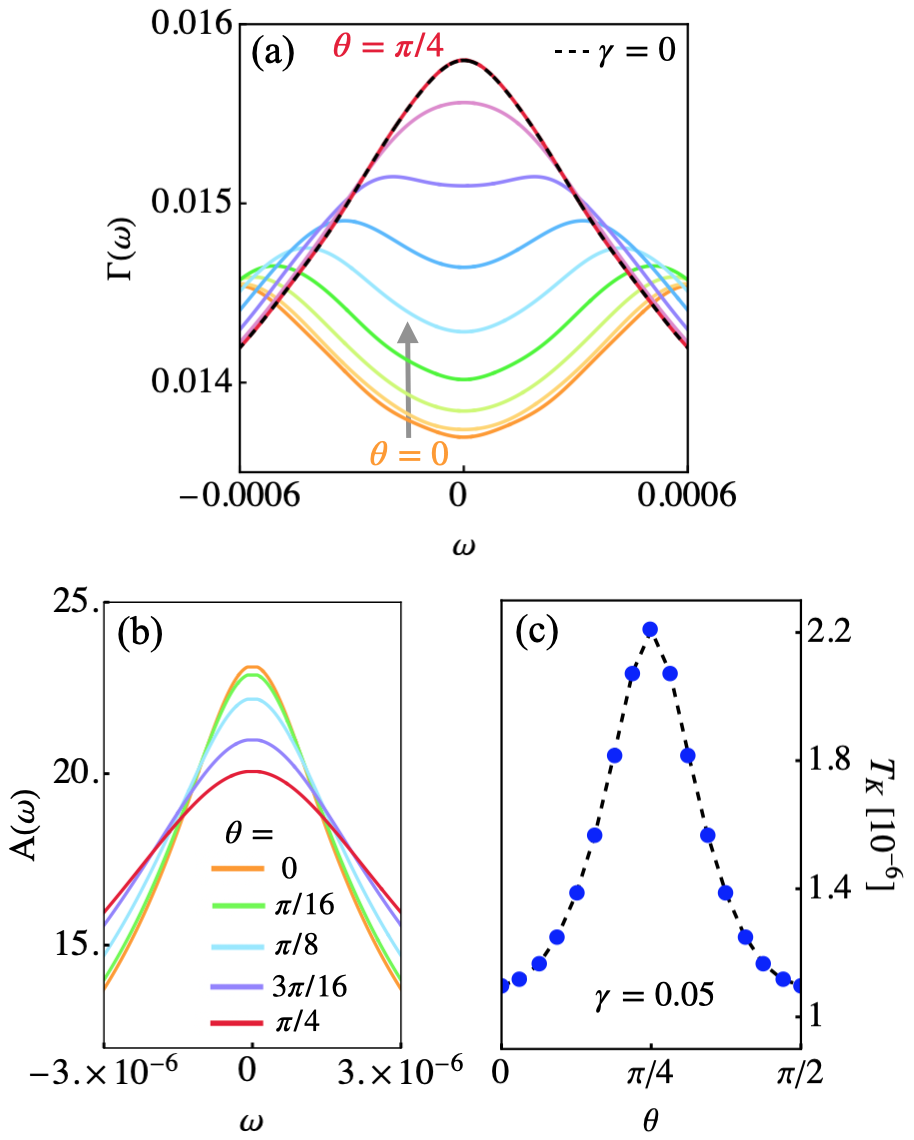} 
	\caption{(a) Closer look at the hybridization-function peak as the spin-orbit contributions (Dresselhaus and Rashba) change by varying $\theta$, for a fixed SOC strength $\gamma=0.05$.
 The light-gray vertical arrow points in the direction of increasing $0 \leq \theta \leq \pi/4$ by steps of $\pi/32$.
 The black dashed curve shows $\Gamma(\omega)$ in the absence of SOC, i.e., $\gamma=0$;
notice that the PSH state ($\theta=\pi/4$) and the absence of SOC are indistinguishable, for $\gamma=0.05$.
(b) impurity's spectral function for some of the $\theta$ values in (a). (c) Kondo temperature $T_K$ variation with the spin-orbit contributions, for $\gamma=0.05$.
\label{fig3}}
\end{figure}

We now examine the combined effect of Rashba and Dresselhaus SOC by adjusting the relative contributions of each through the $\theta$ parameter [see Eq.~\eqref{eq: SOC strength with theta}].
Our analysis shows that moving $\theta$ away from the PSH setup ($\theta = \pi/4$) yields symmetric results to the $\Gamma(\omega)$, whether we move from $\pi/4 \rightarrow 0$ or $\pi/4 \rightarrow \pi/2$.
This indicates that Rashba and Dresselhaus SOC modifies similarly the Kondo effect.
In Fig.~\ref{fig3}(a), we display results only for $0 < \theta < \pi/4$, as the range $\pi/2 > \theta > \pi/4$ exhibits identical behavior.
Due to this symmetry, the hybridization function at $\theta = 0$ matches that at $\theta = \pi/2$, shown in Fig.~\ref{fig2}(d) for $\gamma = 0.05$.
Interestingly, the SOC-induced splitting vanishes at the PSH condition, causing the hybridization function to resemble the case without SOC.

Having obtained the impurity spectral function, we extract the Kondo temperature, for various values of $\theta$.
Theses results are presented in Figs.~\ref{fig3}(b) and~\ref{fig3}(c).
For $A(\omega)$, we show results for some of the $\theta$ values displayed for $\Gamma(\omega)$, while for $T_K$ we show results for $0 < \theta < \pi/2$, clearly illustrating the symmetry around $\pi/4$, as discussed earlier.
In $A(\omega)$,  we observe a similar trend to that in Fig.~\ref{fig2}, i.e., as the peak of the hybridization function increases, the spectral function peak decreases, but its full-width at half maximum increases.
Additionally, the values of $T_K$ are on the order of $\sim 10^{-6}$, which is lower than the estimate given by Haldane’s formula~\cite{Haldane_1978_JPhysC}.
We attribute this discrepancy not only to the choice of $U$, which might be large and extrapolate the validity of the Haldane expression, but mainly because of the non-constant-energy hybridization function.
Notably, as we move from either the pure Rashba or Dresselhaus configurations ($\theta = \pi/2$ or $\theta = 0$) towards the PSH configuration ($\theta = \pi/4$), the Kondo temperature approximately doubles in magnitude, increasing from around $1.1 \times 10^{-6}$ to $2.2 \times 10^{-6}$.

\begin{figure}[h!]
\includegraphics[width=1.0\columnwidth]{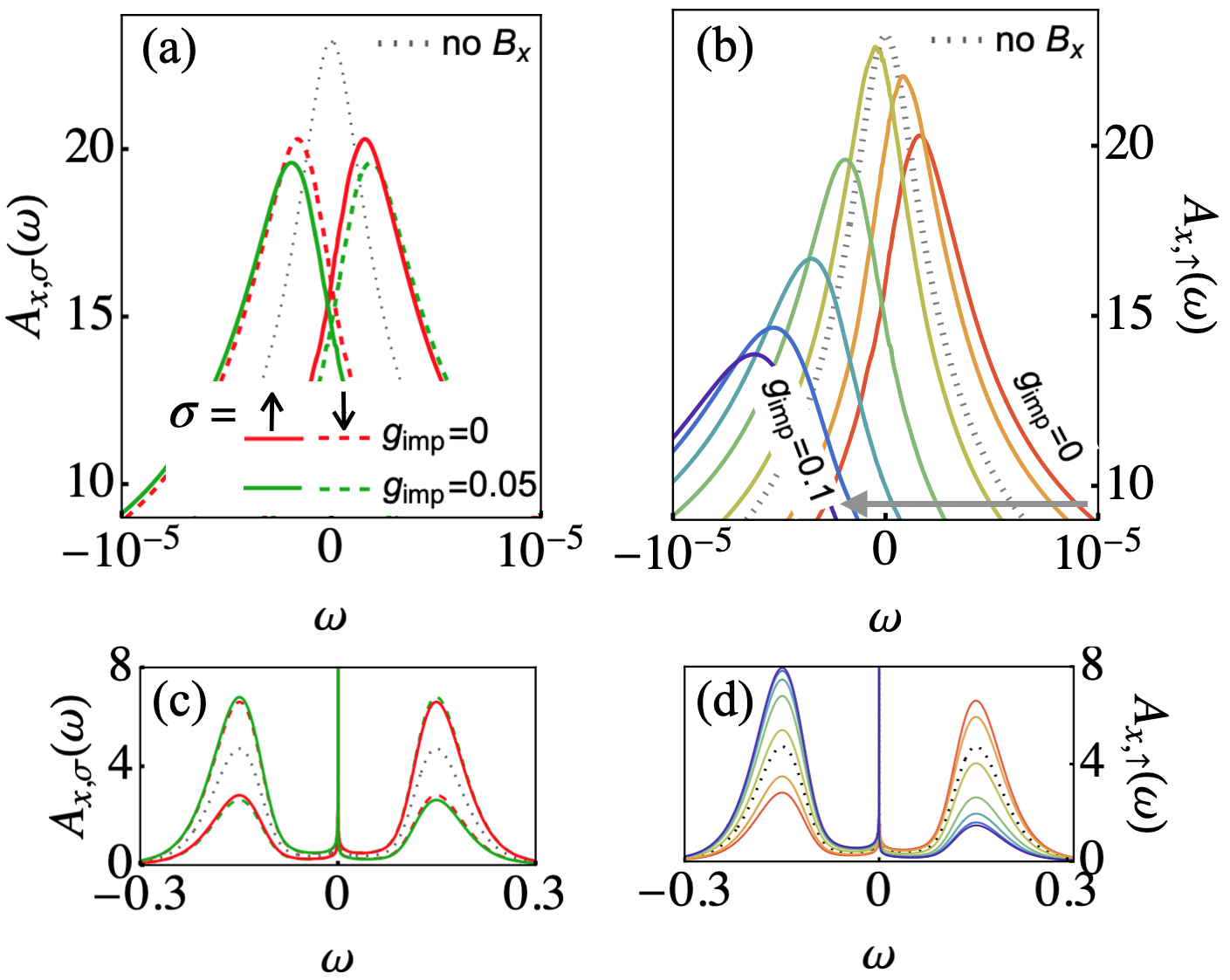}
\caption{
(a) Zoom in on the Kondo peak in the absence ($B_x = 0$) and in the presence ($B_x = 10^{-4}$) of a magnetic field, for the cases in which the impurity g-factor is absent ($g_\text{imp} = 0$) and present ($g_\text{imp} = 0.05$). The SO coupling strength is $\gamma=0.05$ and $\theta = \pi / 2$ (Rashba). The spin-up components (solid lines) are aligned with the magnetic-field direction, while the spin-down components are in opposite direction (dashed line).
(b) shows only the spin-up component of the Kondo peak as the impurity g-factor is increased, the curves follow $g_\text{imp}= 0, 0.01, 0.03, 0.05, 0.07, 0.09, \text{and }0.1$ from right to left.
(c) and (d) are zooms to the Anderson shoulders of (a) and (b), respectively.
}
\label{fig4}
\end{figure}

\subsection{The effect of external magnetic field} \label{sec: mag field}

In this section, we present results for the system under an external magnetic field, $\boldsymbol{B}$.
Important factors come into play: the effect of the magnetic field on the 2DEG spectra, its impact on the impurity, and consequently how these changes influence the Kondo peak.
The application of a magnetic field breaks time-reversal symmetry, leading to a non-zero spin polarization in the conduction band of the 2DEG. 
In the absence of SOC, the magnetic field causes an isotropic splitting of the energy bands via the Zeeman effect, while the presence of SOC induces higher saturation magnetization within the $xy$-plane.
Interestingly, at the PSH condition, we observe band polarization along a direction different from that of the magnetic field. 
A comprehensive discussion of band polarization is provided in the Appendix \ref{app-band-polarization}.
For the impurity, $\boldsymbol{B}$ couples to its spin with a coefficient $g_\text{imp}$, as expressed in Eq.~ \eqref{eq:Hamilt2}.
For simplicity, our analysis below focuses on the case with $\theta = \pi/2$ (Rashba SOC) and $\gamma = 0.05$, with a fixed magnetic field strength of $|\boldsymbol{B}| = 10^{-4}$. 

For a general direction of the external magnetic field, the impurity spectral function acquires a vectorial character, represented as $\boldsymbol{A}_\sigma(\omega) = \left ( A_{x,\sigma}(\omega),A_{y,\sigma}(\omega),A_{z,\sigma}(\omega) \right )$, with $\sigma = \uparrow, \downarrow$.
In Fig.~\ref{fig4}, we show results for $\boldsymbol{B} = B_x \hat{\boldsymbol{x}}$ and the spectral function along this same direction.
Starting with the case $g_\text{imp} = 0$ in Figs. \ref{fig4}(a) and \ref{fig4}(c), we observe a splitting between the spin components $A_{x,\uparrow}(\omega)$ and $A_{x,\downarrow}(\omega)$, along with an asymmetry in the Hubbard peaks at positive and negative energies.
Interestingly, the $\uparrow$-spin component (aligned with the magnetic field direction) appears at a higher energy than the $\downarrow$-spin component.
This effect arises because the external magnetic field polarizes the conduction band, which then couples to the impurity spin in a singlet configuration, aligning the impurity spin opposite to the applied magnetic field.

The scenario shifts when the magnetic field couples directly to the impurity, $g_\text{imp} \neq 0$.
In this case, the $\uparrow$-spin component of the spectral function shows lower energy, which is a result driven by the strength of $g_\text{imp}$. 
Here, there is a competition between the band polarization and the external field in determining the impurity spin orientation.
Figs.~\ref{fig4}(b) and \ref{fig4}(d) illustrate the $\uparrow$-spin component of the spectral function as $g_\text{imp}$ increases.
Notably, $g_\text{imp}$ compensates the effect of the polarized band and, at a certain value, the impurity spectral function resembles the case without an applied magnetic field.
At this point, the spin components $A_{x,\uparrow}(\omega)$ and $A_{x,\downarrow}(\omega)$ become degenerate again, and the Hubbard peaks regain symmetry. This behavior holds independently of the presence of SOC. 
In the Appendix~\ref{app-gimp}, we present similar results for the case $\gamma = 0$.

\begin{figure}[t!]
\includegraphics[width=1.0\columnwidth]{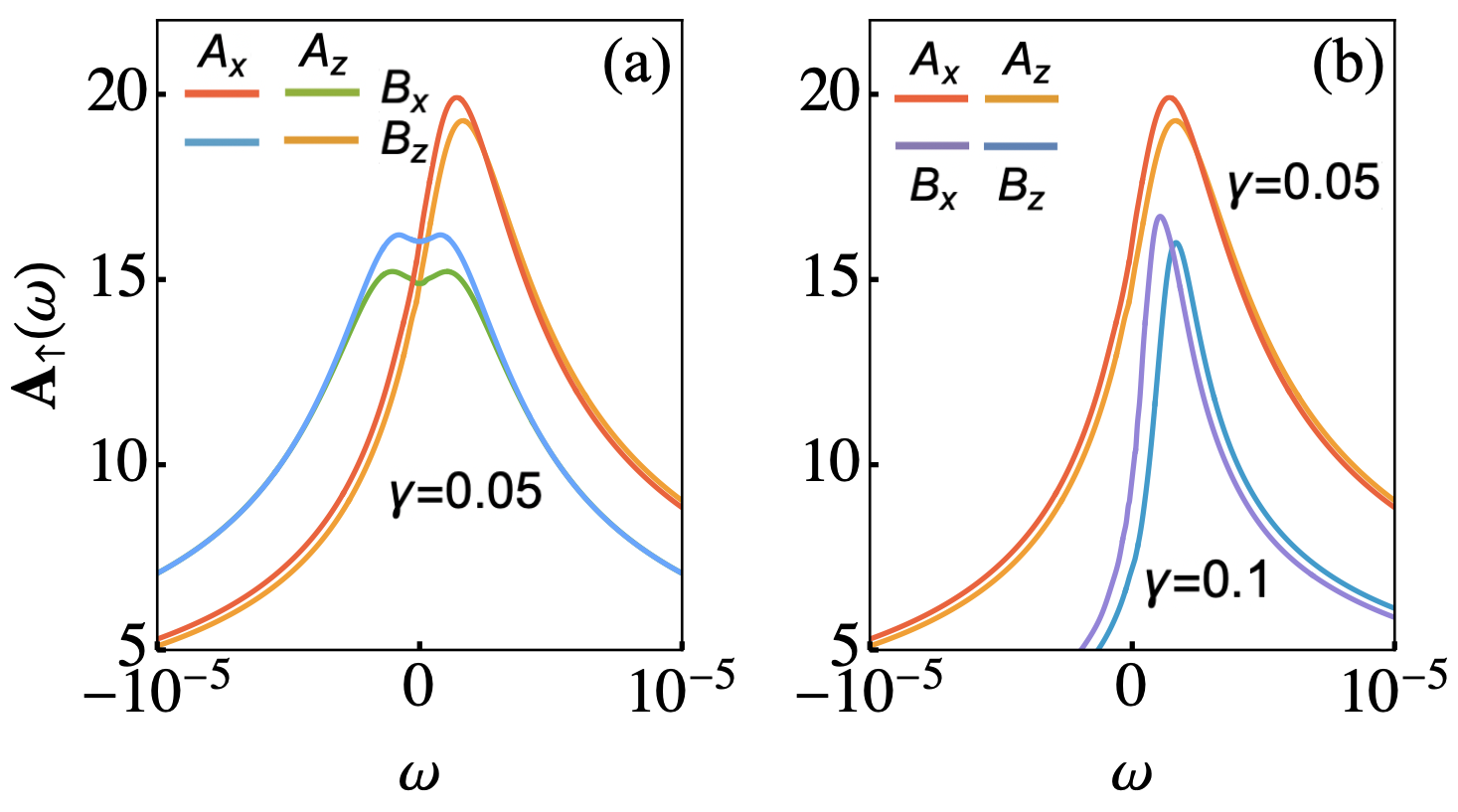}
\caption{
(a) Spectral functions, $A_x(\omega)$ and $A_z(\omega)$, for the cases of external magnetic field oriented within the SO plane ($B_x$) and perpendicularly oriented ($B_z$). 
(b) Spectral functions along the external magnetic field for two values of SO strength ($\gamma$).
In here, we included only Dresselhaus coupling ($\theta = 0$), $|\boldsymbol{B}|=10^{-4}$, and $g_\text{imp}=0$.
}
\label{fig5}
\end{figure}

Now, let us analyze the spectral function components along directions other than 
the magnetic field orientation.
For simplicity, we assume the $g_\text{imp}=0$ and fixed $\theta = 0$ (Dresselhaus SOC).
In Fig.~\ref{fig5}, we present results for the $\uparrow$-spin component, considering in-plane with the SOC and out-of-plane orientations of the magnetic field.
As shown in Appendix~\ref{app-band-polarization}, band saturation polarization is more pronounced when the magnetic field lies within the $xy$-plane.
Similarly, in Fig.~\ref{fig5}(a), contrasting results for the $A_x$ with $B_x$ and $A_z$ with $Bz$ components, we observe a higher Kondo peak when the field aligns within the SOC plane.
An interesting feature appears in the perpendicular components, namely, $A_x$ for $B_z$ and $A_z$ for $B_x$.
Here, we see a noticeable reduction in the Kondo peak, accompanied by a small dip in the peak structure.
This dip arises due to the Zeeman splitting of the conduction band induced by the magnetic field.
In the Appendix~\ref{app-gimp}, we show that a similar structure appears even in the absence of SOC and $g_\text{imp}$.
This feature appears also in the results for an arbitrary magnetic field orientation.

Finally, in Fig.~\ref{fig5}(b), we return to the spectral function components aligned with the external magnetic field, this time for two different SOC strengths.
A comparison of these results to those in Fig.~\ref{fig2} (without a magnetic field), reveals some interesting differences.
Here, as SOC strength increases, the Kondo peak narrows. 
However, unlike the previous case, the peak decreases and it becomes more asymmetric.
Note that the peak is positioned on the positive side of the frequency spectrum due to our choice of $g_\text{imp} = 0$ (see Fig.~\ref{fig4}).
Including $g_\text{imp}$ would primarily shift the peak position.

\begin{figure}[t!]
\includegraphics[width=1.0\columnwidth]{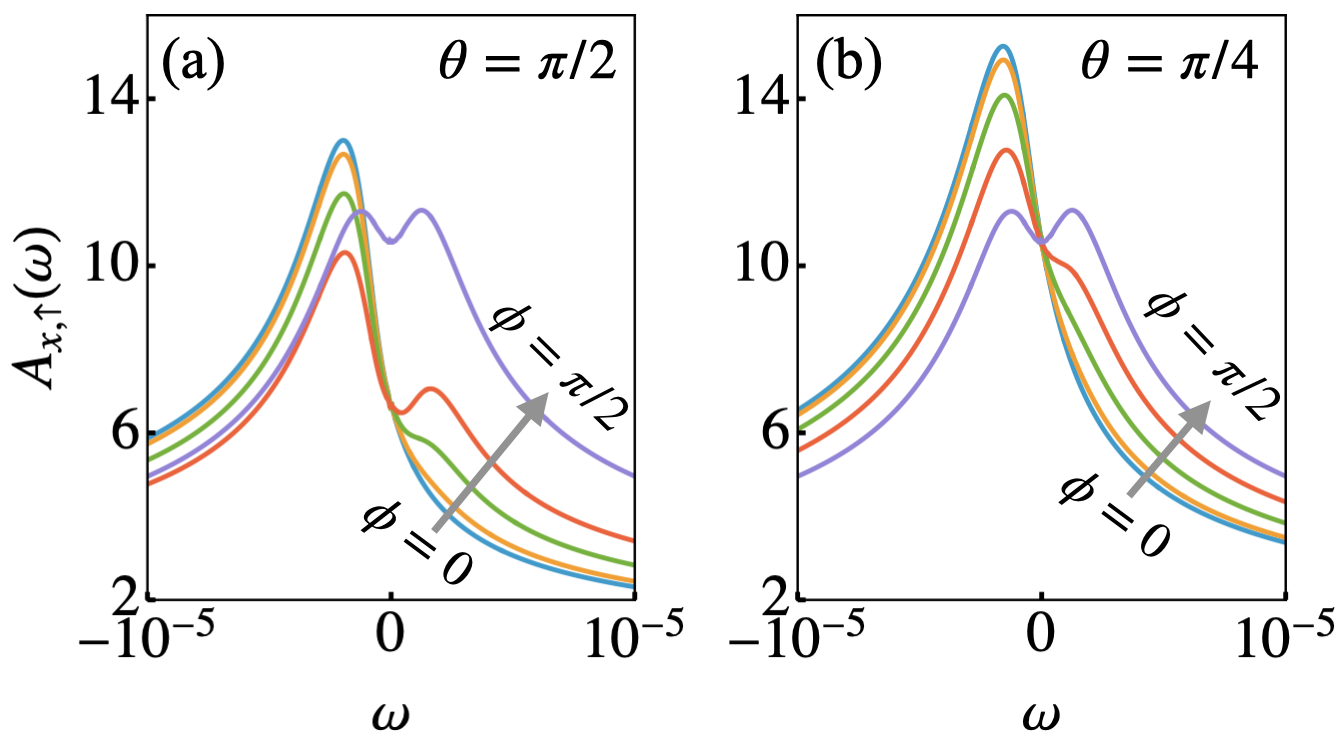}
\caption{
Spectral function, $A_x$, at the (a) Rashba and (b) PSH configuration when varying the magnetic field orientation within the $xy$ plane. For $\phi = 0$ the magnetic field is at the $x$ direction ($B_x$) while for $\phi = \pi/2$ the magnetic field is at the $y$ direction ($B_y$). In here, we used the parameters $|\boldsymbol{B}|=10^{-4}$, and $g_\text{imp}=0.1$
}
\label{fig6}
\end{figure}

\subsection{Varying magnetic field orientation and the PSH condition} 
\label{sec: mag field orientation}

Having discussed the effects of SOC and the magnetic field on the impurity spectral function, we now combine these elements and focus on the unique features of the PSH condition.
For this, we examine various in-plane orientations of the magnetic field with respect to SOC, having defined $\boldsymbol{B} = B_x\cos(\phi) \boldsymbol{x} + B_y \sin(\phi) \boldsymbol{y}$, with a fixed magnitude $|\boldsymbol{B}| = 10^{-4}$.
In Fig.~\ref{fig6}(a), we present $A_{x,\uparrow}(\omega)$ for several values of $\phi$.
Here, $g_\text{imp}$ is included, positioning the peak on the negative side of the frequency spectrum (see Fig.~\ref{fig4} for more details).
Since we are examining the $x$-component of the spectral function, the peak is maximized when the magnetic field aligns with the $x$-direction ($\theta = 0$).
Interestingly, as the magnetic field orientation rotates toward the $y$-direction, the peak diminishes, and a new peak emerges with the opposite frequency sign. 
When the field is fully perpendicular to the $x$-component of $\boldsymbol{A}_{\sigma}(\omega)$, a symmetric double-peak structure appears around $\omega = 0$, similar to the case in Fig.~\ref{fig5}(a), though with a reduced magnitude due to the influence of $g_\text{imp}$.

At last, we examine the results at the PSH condition, as shown in Fig.~\ref{fig6}(b).
For a magnetic field oriented along the $x$-direction ($\phi = 0$), $A_{x,\uparrow}(\omega)$ exhibits a single peak on the negative side of the frequency spectrum, with a height greater than that observed in the Rashba-SOC-only case.
Notably, although Fig.~\ref{fig3}(b) indicated that the PSH configuration results in a higher $T_K$, this enhancement is not observed here when a magnetic field is present.
Similar to the Rashba case, a double-peak structure emerges as the magnetic field rotates toward the $y$-direction.
When the magnetic field is fully perpendicular to the spectral function component, $\boldsymbol{B}\perp \boldsymbol{A}_{\sigma}(\omega)$, the impurity spectral function appears identical for both Rashba and PSH configurations.

\newpage
\section{Conclusions} 
\label{sec: conclusions}

We have presented the interplay between SOC in a 2DEG and magnetic field orientation in the presence of a magnetic impurity using NRG technique.
We analyzed their combined effects on the impurity's spectral function and Kondo temperature.
The break of time-reversal symmetry by the external field raises the vectorial character of $\boldsymbol{\Gamma}(\omega)$ and $\boldsymbol{A}(\omega)$.
By systematically varying the SOC parameter and magnetic field orientation, we uncovered key trends, such as splitting in the hybridization function, increase in the Kondo temperature, and asymmetries in the impurity's spectral function.
In the absence of an external field, the spectral function of the PSH configuration is indistinguishable from that of a system without SOC. When an external field is applied, its orientation induces a smooth change in the spectral function. However, away from the PSH, the spectral function component undergoes an abrupt change when aligned perpendicularly to the magnetic field.
We have also explored different g-factor couplings between the impurity and the external field, which we found to compete with the conduction band spin polarization.
These results emphasize the role of symmetry and competition between SOC and magnetic field effects, contributing valuable insights into the tunability of quantum impurity systems.

\acknowledgments
We thank R. Zitko for enlightening discussions regarding details of the NRG Ljubljana. E.V. acknowledges financial support from the National Council for Scientific and Technological Development (CNPq), Grant No. 311366/2021-0. G.S.D thanks the computer support from LaMCAD/UFG.

\appendix
\renewcommand{\thefigure}{A.\arabic{figure}}
\setcounter{figure}{0}

\section{Explicit form for the $\hat{G}_\text{2DEG}(\omega)$}
\label{app-explicit-G}

In this section we explicitly write the 2DEG Green's function and the non-interacting impurity self-energy. 
First, the Green's function was presented in the main text having the form $\hat{G}_\text{2DEG}(\boldsymbol{k},\omega)=\left[\omega\sigma^{0} -H_\text{2DEG}(\boldsymbol{k})\right]^{-1}$, that can be explicitly written as
\begin{widetext}
\begin{eqnarray}
\hat{G}_{\text{2DEG}}\left(\boldsymbol{k},\omega\right)=D\begin{pmatrix}\left(B_{z}-\epsilon+\omega\right) & \left(B_{x}+k_{y}\alpha+k_{x}\beta\right)-i\left(B_{y}-k_{x}\alpha-k_{y}\beta\right)\\
\left(B_{x}+k_{y}\alpha+k_{x}\beta\right)+i\left(B_{y}-k_{x}\alpha-k_{y}\beta\right) & -\left(B_{z}+\epsilon-\omega\right)
\end{pmatrix}
\label{eq: G2DEG matrix form}
\end{eqnarray}
\end{widetext}
with 
${ D = \left(\det[H_{\text{2DEG}}(\boldsymbol{k})]-2\epsilon\omega+\omega^{2}\right)^{-1} }$, and $H_{\text{2DEG}}(\boldsymbol{k})$ is provided in Eq.~\eqref{eq:Hamilt1compact}.
This Green's function can also be expressed in terms of Pauli matrices as
$\hat{G}_{\text{2DEG}}\left(\boldsymbol{k},\omega\right) = D\left[\left(\omega-\epsilon\right)\sigma_{0}+\boldsymbol{B}\cdot\boldsymbol{\sigma}+\left(k_{y}\alpha+k_{x}\beta\right)\sigma_{x}-\left(k_{x}\alpha+k_{y}\beta\right)\sigma_{y}\right]$.
Finally, the non-interacting self-energy is $\hat{\Sigma}^{(0)}(\boldsymbol{k},\omega)=\hat{V} \hat{G}_\text{2DEG}( \boldsymbol{k},\omega)\hat{V}^{\dagger}$. Because of the choice $\hat{V}=V\sigma^{0}$, the self-energy becomes simply
\begin{equation}
    \hat{\Sigma}^{(0)}(\boldsymbol{k},\omega) = V^2 \hat{G}_\text{2DEG}( \boldsymbol{k},\omega),
\end{equation}
where $\hat{G}_\text{2DEG}( \boldsymbol{k},\omega)$ is given in Eq.~\ref{eq: G2DEG matrix form}.
The hybridization function is then calculated from $\hat{\Sigma}^{(0)}$.
From these results we demonstrate the off-diagonal part of the Green's function, which takes contribution from the spin-orbit coupling and the external field in the $xy$-plane, resulting consequently in the spin-mixing of the hybridization function as well.

\section{Spin resolved component of $G$ along different directions}
\label{app-spin-resolved-G}

Initially, we define the creation operator of a spin $\nicefrac{1}{2}$ state, 
projected along the $z$-direction as 
\begin{eqnarray}
\vert \sigma\rangle = c^{\dagger}_{\sigma}\vert 0\rangle,
\end{eqnarray}
where $\sigma=~~\uparrow, \downarrow$, and the spin projection ($\pm$) along the $x$ and $y$ directions,
\begin{eqnarray}
c^{\dagger}_{x,\pm}\vert 0\rangle\!\!\!&=&\!\!\!\vert\pm\rangle_{x}\! = \!\frac{1}{\sqrt{2}}(\vert\uparrow\rangle \pm \vert\downarrow\rangle)\!=\!\frac{1}{\sqrt{2}}(c^{\dagger}_{\uparrow}\pm c^{\dagger}_{\downarrow})\vert 0\rangle, \\ \nonumber
c^{\dagger}_{y,\pm}\vert 0\rangle\!\!\!&=&\!\!\!\vert\pm\rangle_{x}\! = \!\frac{1}{\sqrt{2}}(\vert\uparrow\rangle \pm i \vert\downarrow\rangle)\!=\!\frac{1}{\sqrt{2}}(c^{\dagger}_{\uparrow}\pm ic^{\dagger}_{\downarrow})\vert 0\rangle. 
\end{eqnarray}
Thus, in Zubarev's notation, the $2 \times 2$ spin-resolved Green's function can be written 
in a compact form, $G_{\sigma\sigma^{\prime}}(\omega)=
\langle \langle c_{\sigma};c^{\dagger}_{\sigma^{\prime}} \rangle \rangle_\omega$, 
from where we can compute the spin-resolved projection in a given 
direction as
\begin{subequations}
\begin{eqnarray}
\label{Gss}
G_{x,\pm}&=& \dlangle c_{x,\pm};c^{\dagger}_{x,\pm}\drangle=\frac{1}{2}\dlangle (c_{\uparrow}\pm c_{\downarrow});(c^{\dagger}_{\uparrow}\pm c^{\dagger}_{\downarrow})\drangle, \qquad\\ 
G_{y,\pm}&=& \dlangle c_{y,\pm};c^{\dagger}_{y,\pm}\drangle=\frac{1}{2}\dlangle (c_{\uparrow}\mp ic_{\downarrow});(c^{\dagger}_{\uparrow}\pm ic^{\dagger}_{\downarrow})\drangle, \qquad \\
G_{z,\uparrow}&=& \dlangle c_{\uparrow};c^{\dagger}_{\uparrow}\drangle, \\ 
G_{z,\downarrow}&=& \dlangle c_{\downarrow};c^{\dagger}_{\downarrow}\drangle.
\end{eqnarray}
\end{subequations}
From Eq.~\ref{Gss}, we can calculate the spin-resolved density of states,
\begin{eqnarray}
\rho_{(i=x,y),\pm}&=& -\frac{1}{\pi}{\rm Im}[G_{i,\pm}],\\\nonumber
\rho_{z,\sigma=\uparrow,\downarrow}&=& -\frac{1}{\pi}{\rm Im}[G_{z,\sigma}].
\end{eqnarray}

\section{Method details}
\label{app-method-details}

The NRG method consists of a logarithmic discretization of the hybridization function in energies given by $\varepsilon_N=(1/2)(1+\Lambda^{-1})\Lambda^{-(N-1)/2}$, rendering energy intervals that decreases as $\Lambda^{-N}$ when $N$ increases. Strictly speaking, the Fermi energy is reached  when $N\rightarrow \infty$. Here, $\Lambda > 1$ is generically called the discretization parameter. The discretized version of the Hamiltonian is further written in a  tridiagonalization form, which results in a one-dimensional tight-binding Hamiltonian (a.k.a. Wilson chain Hamiltonian). Once the tridiagonal Hamiltonian is obtained an iterative diagonalization  is performed with a proper truncation of the Hilbert space~\cite{Wilson1975}. 
The nontrivial energy dependence of the hybridization function treated here requires an improved discretization scheme to reproduce with high resolution the spectral functions and reduce the numeric inaccuracies in the NRG calculations. 
We then adopt the adaptive ``z-averaging" 
scheme~\cite{PhysRevB.79.085106}, as implemented in the 
NRG Ljubljana open source code~\cite{zitko_rok}. 
We selected the parameters $U=0.3$ and $V=0.05$ (this leading to $\Gamma(0) = \pi V^2 \rho(\varepsilon_F) \sim 0.015$) to achieve a Kondo temperature ($T_K$) on the order of $\sim 10^{-6}$, estimated using Haldane's formula~\cite{PhysRevLett.40.416,PhysRevLett.40.911.2,Haldane_1978_JPhysC},
\begin{eqnarray}
    T_K \sim \sqrt{ \frac{2 U \Gamma (0)}{\pi} } \exp \left [ \frac{\epsilon (\epsilon+U)}{2 U \Gamma (0) / \pi} \right ].
\end{eqnarray}

In Fig. 3(c) in the main text we extracted the Kondo temperature using the Wilson relation $\chi_\text{imp} = 0.413 / 4 T_K$~\cite{Bulla2008}, where $\chi_\text{imp}$ is the impurity's susceptibility.

\section{Band Polarization}
\label{app-band-polarization}

To understand how the 2DEG band \emph{polarizes} as a function 
of the external magnetic field, we have numerically calculated the spin expectation values 
$\langle S_{i}\rangle=\Sigma_{k}\langle k \vert S_{i} \vert k\rangle$ ($i=x,y,z$), 
for the external magnetic field $\mathbf{B}$ along the coordinate axes, where $\vert k \rangle$ 
represents the Bloch states obtained through the diagonalization of $H_{\rm 2DEG}$.

\begin{figure}[t!]
\includegraphics[width=0.85\columnwidth]{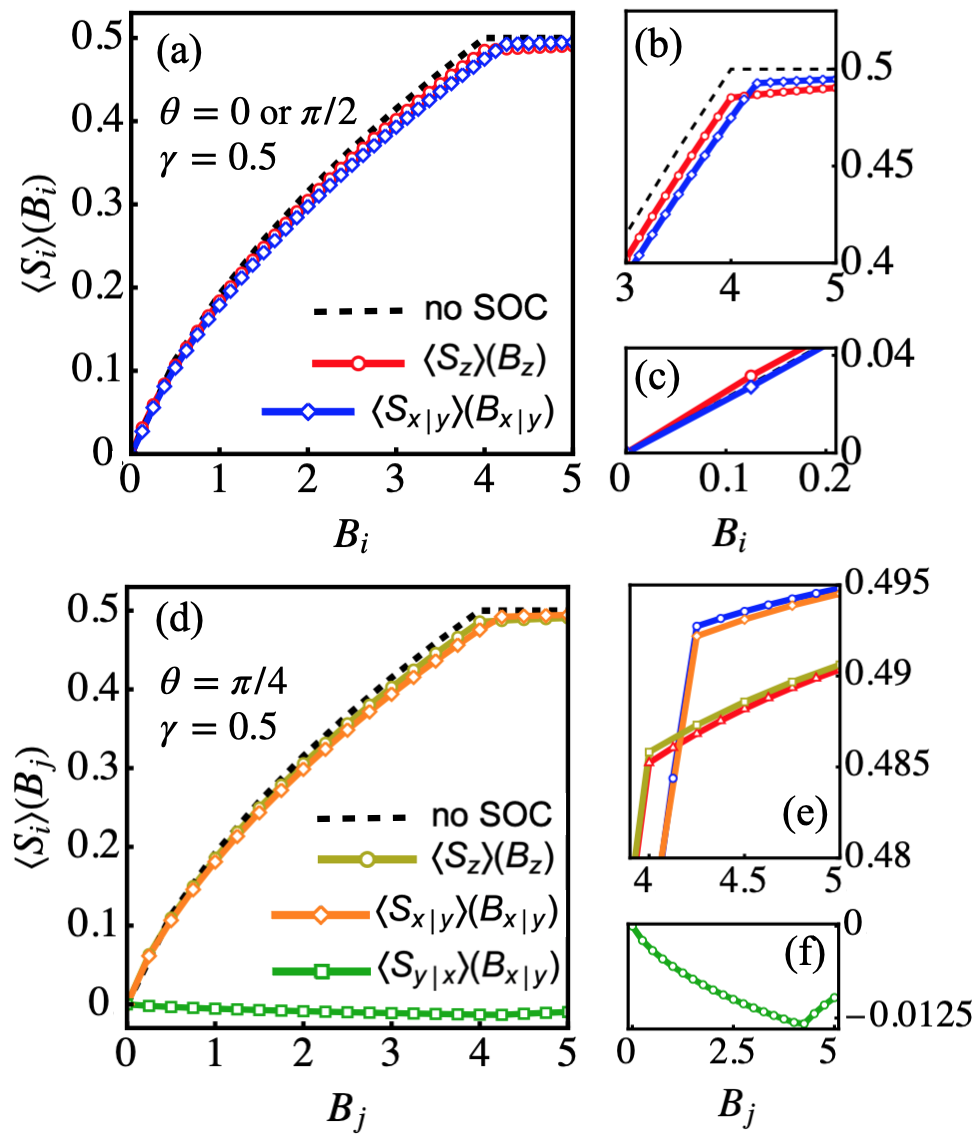}
\caption{Spin polarization of the conduction band $\langle S_{i}\rangle (B_j)$ ($i,j=x,y,z$) resulting from an external magnetic field applied along different axes. 
(a) shows the spin polarization along the magnetic field in the absence (no SOC) and in the presence ($\gamma = 0.5$) of spin-orbit coupling, for the limiting cases of only Dresselhaus ($\theta = 0$) or Rashba ($\theta = \pi/2$) coupling;
notice that, $\langle S_x \rangle (B_x) = \langle S_y \rangle (B_y)$ (red curve) differs from $\langle S_z \rangle(B_z)$ (blue curve), as shown in the zoom in on high field (a1) and low field (a2) values. 
(b) shows the spin polarization at the PSH state ($\theta = \pi/4$), in which an additional crossed diamagnetic polarization (e.g., $\langle S_x \rangle (B_y)$) appears when the magnetic field is in the $xy$ plane, see (b2) for a zoom in.
A better comparison between the curves in (a) and (b) can be seen in (c) for large values of magnetic field, in which we show the quantities $\langle S_z \rangle (B_z)$ and $\langle S_{x|y} \rangle (B_{x|y})$ following the same legends as in (a) and (b).
\label{figA1}}
\end{figure}

Figure~\ref{figA1} shows the band polarization results $\langle S_i \rangle$ when the 
external magnetic field is applied along the three coordinate axes. 
$\langle S_{i}\rangle(B_j)$ represents the expectation value of the spin along $i$-direction, in the presence of a magnetic field applied at the $j$-direction.
For zero-SOC [dashed black curve in panel (a)], the system is spatially 
isotropic and an external magnetic field splits isotropically 
the energy bands due to the Zeeman field (as long as the $g$-factor 
is isotropic~\cite{PhysRev.174.823}). As 
the external magnetic field lifts the time reversal symmetry, 
a non-zero spin polarization is observed. That polarization appears 
only in the direction of the applied field and it is independent of 
the axis along which the external magnetic field is applied [black 
dashed line in Fig.~\ref{figA1}(a)]. 
In Fig.~\ref{figA1}(a), we present results for $\alpha=0.25$ 
and $\beta=0$ (or $\beta=0.25$ and $\alpha=0$), i.e., Rashba 
only (or Dresselhaus only). As for the zero-SOC case, there 
is polarization only along the direction of the applied magnetic 
field, but now, due to the presence of an (in-plane) momentum-dependent 
effective $\mathbf{B_{SO}(\mathbf{k})}$, part of the isotropy is lost. 
Indeed, $\langle S_x \rangle = \langle S_y \rangle \ne \langle S_z \rangle$ 
(black, blue, and red circles, respectively), with a higher in-plane 
saturation polarization than out-of-plane, with both smaller than 
the zero-SOC saturation value. It is worth noting that, for 
the small values of external field that will be used for the 
probing of the Kondo state [see inset in panel (a)], the 
out-of-plane polarization is higher than the in-plane 
polarization. 

Figure~\ref{figA1}(b) shows the results for finite SOC, 
but in the PSH regime ($\alpha=\beta=0.5/\sqrt{2}$). 
The overall result is similar to the one for the Rashba-only and 
Dresselhaus-only results in panel (a), with an interesting difference: 
an additional (small) crossed diamagnetic polarization (negative 
$\langle S_x \rangle$ for applied $B_y$, blue squares, and 
negative $\langle S_y \rangle$ for applied $B_x$, black squares) 
is observed for in-plane external magnetic field. A zoom of 
these diamagnetic contributions is shown in the inset. 

It is important to stress that a magnetic field perpendicular 
to the 2DEG $xy$-plane fully quantizes the motion of in-plane electrons 
(holes) into Landau levels~\cite{PhysRevB.72.085342}. However, in 
our numerical calculations we have neglected the orbital effects 
of $\mathbf{B}$, assuming instead a system where the ratio 
of Zeeman energy ($\varepsilon_{Z}$) and the cyclotron frequency 
($\hbar\omega_{c}$) is high~\cite{PhysRevB.73.235306}. In addition, 
for the Kondo physics, we will be interested in the regime of 
low magnetic field, otherwise the system (host and impurity) 
will be fully polarized, which tends to suppress the Kondo 
physics.

\begin{figure}[t!]
\includegraphics[width=1.0\columnwidth]{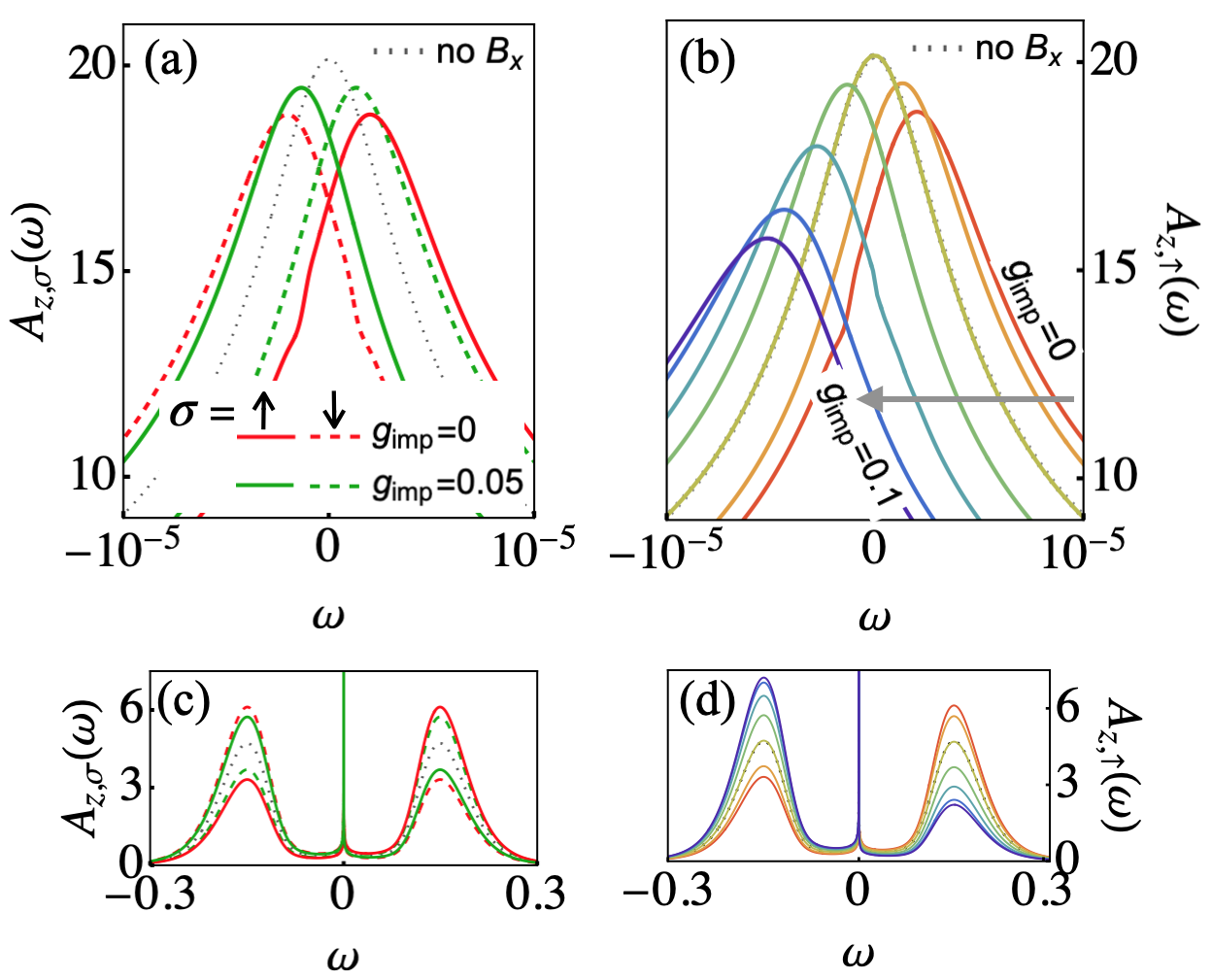}
\caption{
(a) shows a zoom in to the Kondo peak in the absence ($B_z = 0$) and in the presence ($B_z = 10^{-4}$) of a magnetic field, for the cases in which the impurity g-factor is absent ($g_\text{imp} = 0$) and present ($g_\text{imp} = 0.05$). There is no SO ($\gamma=0$). The spin-up components (solid lines) are aligned with the magnetic-field direction, while the spin-down components are in opposite direction (dashed line).
(b) shows only the spin-up component of the Kondo peak as the impurity g-factor is increased, the curves follow $g_\text{imp}= 0, 0.01, 0.03, 0.05, 0.07, 0.09, \text{and }0.1$ from right to left.
(c) and (d) are a zoom in to the Anderson shoulders of (a) and (b), respectively.
}
\label{figA2}
\end{figure}

\begin{figure}[t!]
\includegraphics[width=1.0\columnwidth]{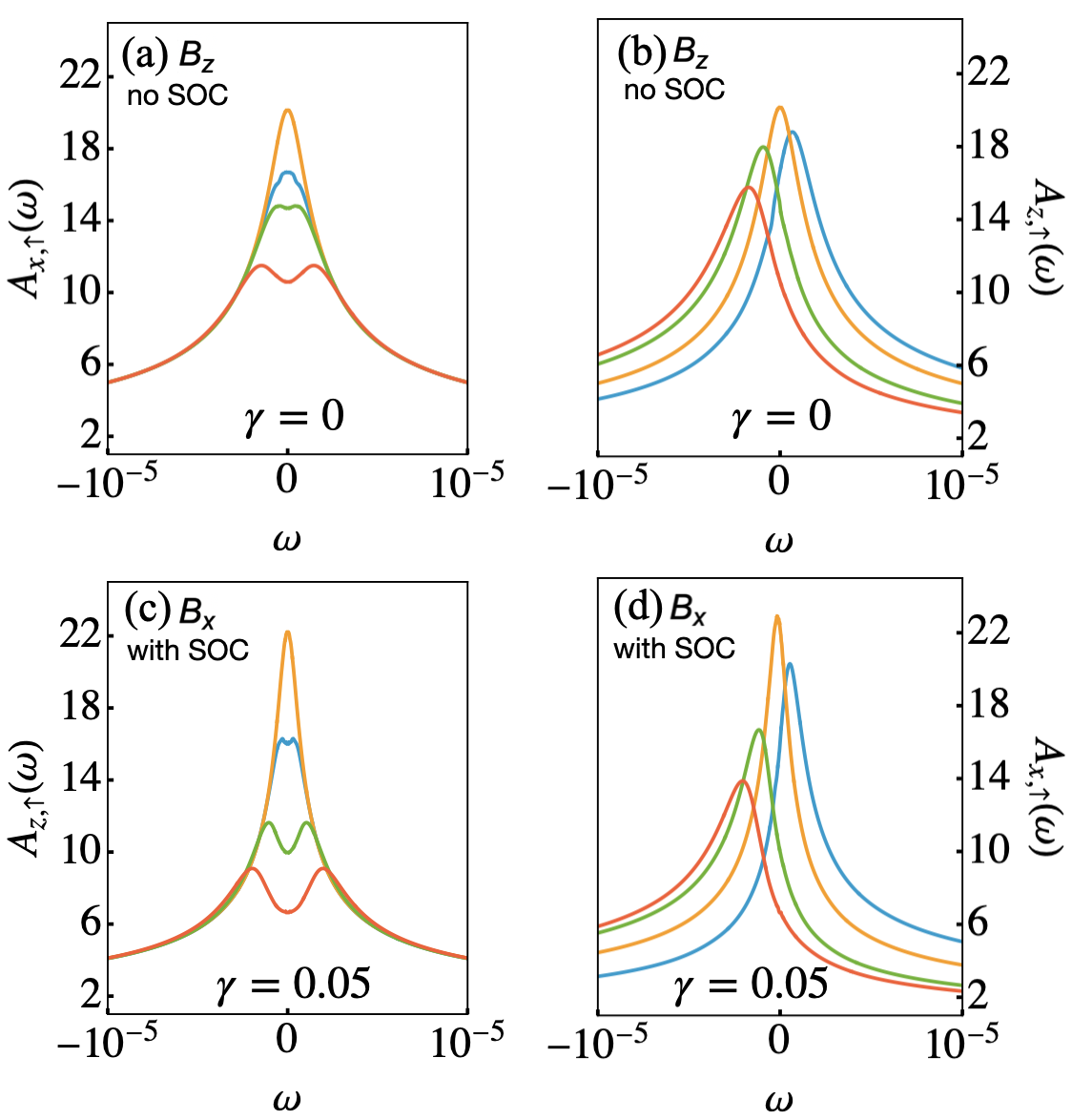}
\caption{
(a) $x$-component and (b) $z$-component of $\boldsymbol{A}_\uparrow (\omega)$ with $\boldsymbol{B}$ oriented along the $z$-direction,
for the case without SOC, shown for several values of coupling $g_\text{imp}$.
(c) $z$-component and (d) $x$-component of $\boldsymbol{A}_\uparrow (\omega)$ with $\boldsymbol{B}$ oriented along the $x$-direction (within the SO-plane), in the presence of SOC.
}
\label{figA3}
\end{figure}

\section{Effect of $g_\text{imp}$}
\label{app-gimp}

In the main text, we have shown how the impurity's spectral function shapes itself for different values of coupling to the external field, $g_\text{imp}$, see Fig.~\ref{fig5}.
In that case we positioned the external field within the SO plane and looked at the spectral function oriented along the field.
Here, in Fig.~\ref{figA2} we eliminate the SO component, such that the 2DEG is isotropic  along all directions, and we are able to isolate the competition effect between the external field at the impurity and the impurity's coupling to the polarized band.
The results are similar to the main text, in which the main different being the scale of the spectral function overall, that is slightly smaller without SOC.

Moreover, in Fig.~\ref{figA3} we analyze the $\boldsymbol{A}_\uparrow(\omega)$ components, parallel and perpendicular to $\boldsymbol{B}$, in the presence and absence of SOC.
For the spectral function components along the magnetic field, the behavior is similar to that discussed in Fig.~\ref{figA2}.
For the perpendicular component, Figs.~\ref{figA2}(a) and \ref{figA2}(c), we observe how the double peak structure changes as we vary $g_\text{imp}$.
The double peak has origin in the time-reversal symmetry break.
As discussed in the main text, there is a competition between the coupling to the external field and the conduction band polarization.
Interestingly, in these figures, we see a cancellation of those two couplings for $g_\text{imp} = 0.03$ (therefore, no double peak structure).

\section{Representative parameters}
\label{app-representative-parameters}

In Table \ref{tbl:parameters} we show the representative parameters used in the NRG Ljubljana. The conduction band discretization is characterized by the parameter $\Lambda$ and the total number of discrete points $N_\text{max}$. 
The bath logarithmic discretization is calculated as
\begin{equation}
    \omega_N = \frac{1 - \Lambda^{-1}}{2} \Lambda^{-(N-1)/2 + 1 - z},
\end{equation}
where $N = 1,\ldots,N_\text{max}$, and the variable $z$ consists in performing different NRG runs then averaging the results to correct the systematic errors. The input file receives the maximum value for the averaging, defined as $N_z=8$.

\begin{table}[t!]
\hfill{}
\begin{tabular}{|c|c|c|c|c|}
\hline 
\textbf{parameter} & \textbf{value} &  & \textbf{parameter} & \textbf{value}\tabularnewline
\hline 
\hline 
$U$ & $0.3$ &  & broaden\_min & $10^{-7}$\tabularnewline
\hline 
$\epsilon$ & $-0.15$ &  & $\omega_n$ & $1.01^n$\tabularnewline
\hline 
$\Lambda$ & $2.25$ &  & $\alpha$ & $0.3$\tabularnewline
\hline 
$N_\text{max}$ & $60$ &  & broaden\_gamma & $0.02$\tabularnewline
\hline 
keep & $2000$ &  & bins & $300$\tabularnewline
\hline 
$N_{z}$ & $8$ &  & $T$ & $10^{-10}$\tabularnewline
\hline 
\end{tabular}
\hfill{}
\caption{Parameters used in the NRG Ljubljana.}
\label{tbl:parameters}
\end{table}

The parameters $\alpha$ are related to the broadening of the spectral-function delta peaks. Within the input file for NRG, $\alpha$ appears as broaden\_alpha parameter, and controls the width of the log-Gaussian broadening kernel.
The variable keep is used to control the truncation of states during the NRG iteration, namely, it sets the absolute upper limit to the number of states kept.
The $\omega_n$ defines the mesh for the broadened spectral functions, where $n=0,\ldots,n_\text{max}$ is defined so that $\omega_{n_\text{max}} > \text{broaden\_min}$. In turn, $\omega_n$ is calculated from two parameters in the input file as $\omega_n = \text{broaden\_max} \times \text{broaden\_ratio}^n$, that we have set $\text{broaden\_max} = 1$ and $\text{broaden\_ratio} = 1.01$.

\newpage

\bibliography{refs}
\end{document}